\documentclass[prl,twocolumn,superscriptaddress,showpacs]{revtex4-1}

\usepackage{amsmath,amssymb,graphicx,color,bm,soul}
\usepackage[usenames,dvipsnames]{xcolor}

\begin{document}

\author{T.~C.~H.~Liew}
\affiliation{Division of Physics and Applied Physics, Nanyang Technological University 637371, Singapore}

\author{Y.~G.~Rubo}
\affiliation{Instituto de Energ\'ias Renovables, Universidad Nacional Aut\'onoma de M\'exico, Temixco, Morelos, 62580, Mexico}

\author{A.~S.~Sheremet}
\affiliation{Russian Quantum Center, Novaya 100, 143025 Skolkovo, Moscow Region, Russia}
\affiliation{Department of Theoretical Physics, Peter the Great St.-Petersburg Polytechnic University, 195251, St.-Petersburg, Russia}

\author{S.~De Liberato}
\affiliation{School of Physics and Astronomy, University of Southampton, Highfield, Southampton, SO17 1BJ, UK}

\author{I.~A.~Shelykh}
\affiliation{Division of Physics and Applied Physics, Nanyang Technological University 637371, Singapore}
\affiliation{Science Institute, University of Iceland, Dunhagi-3, IS-107, Reykjavik, Iceland}
\affiliation{National Research University for Information Technology, Mechanics and Optics (ITMO), 197101 St.-Petersburg, Russia}

\author{F.~P.~Laussy}
\affiliation{Russian Quantum Center, Novaya 100, 143025 Skolkovo, Moscow Region, Russia}
\affiliation{Departamento de F\'isica Te\'orica de la Materia Condensada and Condensed Matter Physics Center (IFIMAC), Universidad Aut\'onoma de Madrid, E-28049, Spain}

\author{A.~V.~Kavokin}
\affiliation{Russian Quantum Center, Novaya 100, 143025 Skolkovo, Moscow Region, Russia}
\affiliation{School of Physics and Astronomy, University of Southampton, Highfield, Southampton, SO17 1BJ, UK}
\affiliation{Spin Optics Laboratory, St.-Petersburg State University, 198504 Peterhof,
St.-Petersburg, Russia}

\title{Quantum Statistics of Bosonic Cascades}
\date{\today}

\begin{abstract}
Bosonic cascades formed by lattices of equidistant energy levels sustaining
radiative transitions between nearest layers are promising for the
generation of coherent terahertz radiation. We show how, also for the light
emitted by the condensates in the visible range, they introduce new regimes
of emission. Namely, the quantum statistics of bosonic cascades exhibit
super-bunching plateaus. This demonstrates further potentialities of bosonic
cascade lasers for the engineering of quantum properties of light useful for
imaging applications.
\end{abstract}


%78.67.-n 	Optical properties of low-dimensional, mesoscopic, and nanoscale materials and structures (for magnetic properties of nanostructures, see 75.75.-c; for electronic transport in nanoscale structures, see 73.63.-b; for mechanical properties of nanoscale systems, see 62.25.-g)
%71.35.-y 	Excitons and related phenomena
%42.50.Ar 	Photon statistics and coherence theory
%78.45.+h 	Stimulated emission (see also 42.55.-f Lasers)

%03.75.Mn 	Multicomponent condensates; spinor condensates
%78.67.Pt 	Multilayers; superlattices; photonic structures; metamaterials (see also %81.05.Xj, Metamaterials for chiral, bianisotropic and other complex media)
%78.66.Fd 	III-V semiconductors
%71.36.+c 	Polaritons (including photon-phonon and photon-magnon interactions)

\maketitle

\emph{Introduction.---}Terahertz frequency radiation is a valuable
resource for a number of imaging applications in medicine, security
screening, and different scientific
disciplines~\cite{tonouchi07a}. For this reason a number of
solid-state systems have been considered as terahertz sources, where
one typically aims to convert an optical frequency photon into a
terahertz photon. Even if materials with appropriate transitions can
be found, such a procedure is inherently inefficient as the majority
of the energy of each optical frequency photon is given up in favour
of generating a low-energy terahertz photon. Quantum cascade lasers
circumvent this problem, where an optical quantum of energy can be
used to generate multiple terahertz
photons~\cite{Kohler2002,Liu2013}. While fermionic (electronic)
cascades have been realised over 20 years ago~\cite{Faist1994},
bosonic cascades were only recently designed theoretically based on
excitonic transitions in parabolic quantum wells, possibly being
placed inside semiconductor microcavities~\cite{Liew2013}.  When using
also a THz cavity, bosonic cascades offer the prospects of double
stimulation of the emission: by the final exciton state
population~\cite{VCSEL, Simone} and by the terahertz cavity mode
population~\cite{kaliteevski14a,pervishko}.

Stimulated transitions among macroscopically occupied bosonic quantum
states are in the heart of bosonic cascade lasers (BCLs), which also
act as polariton lasers spontaneously emitting coherent light in the
optical frequency range~\cite{BosonicLasers}. Polariton lasers bring
the advantage of ultra-low threshold power~\cite{Christopoulos, Hoefling,
  Bhattachariya} and as versatile all-optical
platforms~\cite{sanvitto11a} they are expected to constitute building
blocks of future optical integrated circuits~\cite{Circuits}, logic
elements~\cite{bistability} or polarisation
switches~\cite{switches}. The quantum optical properties of polariton
lasers have been studied theoretically~\cite{laussy04a} and
experimentally~\cite{richard}. The quantum theory of polariton based
terahertz sources has been addressed recently~\cite{delValle2011}. The
quantum physics of bosonic cascades is yet to be explored.

In this work, we open a new dimension for BCLs by showing that their
emission produces a marked superbunching, thanks to their specificity of
making coexist multiple macroscopic coherent states. This makes for a
difficult problem to describe quantum mechanically, that we can tackle here
with stochastic quantum Langevin equations. This theory of bosonic cascades
to describe their statistical properties not only confirms their departures
from other mechanisms of lasing~\cite{paschotta_book08a}, it also suggests
that they could power imaging applications with resolutions much higher than
ever conceived before~\cite{Gong2015}, thanks to combinations of several
assets to go beyond the diffraction limit~\cite{Hong2012,Grujic2013}.

% Remarkably, under a coherent excitation, we see a
% pronounced parity dependent superbunching of bosons, defined by a
% second-order correlation function greatly exceeding $2$.  In
% addition, the second-order correlation function exhibits
% oscillations in time before reaching a terahertz lasing coherent
% state at long times~\cite{delValle2011}.

\begin{figure}[tbp]
\includegraphics[width=.9\linewidth]{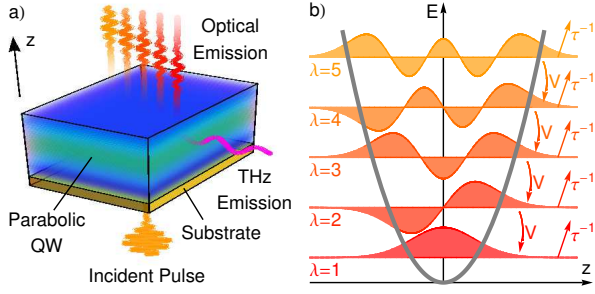}
\caption{Illustration of a bosonic cascade. a) Structure schematic: a
  parabolic quantum well is excited with an optical pulse and emits
  light at a range of frequencies. b) Energy level diagram: the
  parabolic trapping potential engineers equidistant energy levels and
  transitions occur between neighbouring levels with transition
  element $V$. Bosons in each level can decay radiatively, with
  characteristic decay time $\protect\tau$.}
\label{fig:Scheme}
\end{figure}

\emph{Theory.---}We consider a bosonic cascade with $N$ energy levels
labelled from $\lambda =1$ to $N$, as illustrated in
Fig.~\ref{fig:Scheme}.  The parabolic quantum well makes the levels
equidistant in energy~\cite{Faist}%
. We account for the dominating transitions which are between
neighbouring excitonic levels $\lambda +1$ and $\lambda $ and assume
for simplicity that all levels decay radiatively with the same rate,
$\tau ^{-1}$. We assume that the occupation of the terahertz mode fed
by the interlevel transitions in the cascade is zero, i.e., there is
no terahertz cavity in our model structure. This corresponds to the
original version of BCL\cite{Liew2013}, where the stimulation of
terahertz radiation is achieved due to the macroscopic exciton
populations in the bosonic cascade.  This simpler configuration
already manifests the non-trivial quantum effects of bosonic
cascades. The system is thus described by the density matrix
$\hat{\rho}$ subject to a quantum Liouville equation:
\begin{multline}
\frac{d\hat{\rho}}{dt}=\frac{1}{2\tau }\sum_{\lambda =1}^{N}\left( \big[\hat{%
a}_{\lambda }^{\vphantom{\dagger}},\hat{\rho}\hat{a}_{\lambda }^{\dagger }%
\big]+\big[\hat{a}_{\lambda }^{\vphantom{\dagger}}\hat{\rho},\hat{a}%
_{\lambda }^{\dagger }\big]\right)  \\
+\frac{V}{2}\sum_{\lambda =1}^{N-1}\left( \big[\hat{a}_{\lambda }^{\dagger }%
\hat{a}_{\lambda +1}^{\vphantom{\dagger}},\hat{\rho}\hat{a}_{\lambda
+1}^{\dagger }\hat{a}_{\lambda }^{\vphantom{\dagger}}\big]+\big[\hat{a}%
_{\lambda }^{\dagger }\hat{a}_{\lambda +1}^{\vphantom{\dagger}}\hat{\rho},%
\hat{a}_{\lambda +1}^{\dagger }\hat{a}_{\lambda }^{\vphantom{\dagger}}\big]%
\right) \,,  \label{eq:master}
\end{multline}
where $\hat{a}_{\lambda }$ annihilates a boson from level $\lambda $ and $V$
is the scattering rate~\cite{Vindependenceonlambda}. Dotting this equation
with Fock states and neglecting correlations between levels in the
Born-Markov approximation leads to the Boltzmann equations, which describe
well the average populations of each state and thereby provide the
mean-field theory of bosonic cascades~\cite{Liew2013}. To go beyond this
semi-classical description, and in particular to compute the photon
statistics, correlations between states must be retained. A brute-force
approach, however, is restricted to few levels and each with a modest
occupancy. To access the most general cases of bosonic cascades, with
macroscopic occupation of the states, we therefore expand the density matrix
on the natural basis for this problem, that of coherent states~\cite%
{Drummond1980}:
\begin{equation}
\hat{\rho}=\int \mathcal{P}(\alpha _{1},\beta _{1},\ldots ,\alpha _{N},\beta
_{N})\hat{\Lambda}(\alpha _{1},\beta _{1},\ldots ,\alpha _{N},\beta
_{N})d\mu \,,  \label{eq:PositiveP}
\end{equation}%
where:
\begin{equation}
\hat{\Lambda}(\alpha _{1},\beta _{1},\ldots ,\alpha _{N},\beta _{N})=\frac{%
|\alpha _{1},\ldots ,\alpha _{N}\rangle \langle \beta _{1}^{\ast },\ldots
,\beta _{N}^{\ast }|}{\langle \beta _{1}^{\ast },\ldots ,\beta _{N}^{\ast
}|\alpha _{1},\ldots ,\alpha _{N}\rangle }\,,
\end{equation}%
with $\mathcal{P}$ the positive-P distribution, which differs from the
Glauber-Sudarshan distribution in allowing for non-diagonal coherent state
projectors. The complex numbers $\alpha _{\lambda }$ and $\beta _{\lambda }$
are independent variables covering the whole complex plane. The integration
measure is $d\mu =d^{2}\alpha _{1}d^{2}\beta _{1}\ldots d^{2}\alpha
_{N}d^{2}\beta _{N}$ and for ease of notation we write $\vec{\alpha}=(\alpha
_{1},\beta _{1},\ldots ,\alpha _{N},\beta _{N})$. $\vec{\alpha}_{n}$ will be
used to refer to an element of this vector (of length $2N$), while the
notations $\alpha _{\lambda }$ and $\beta _{\lambda }$ will still be used
where $\lambda $ is the level index. Writing the density matrix in terms of
the positive-P distribution (Eq.~\ref{eq:PositiveP}) transforms the master
equation Eq.~(\ref{eq:master}) into a Fokker-Planck equation:
\begin{multline}
\frac{\partial \mathcal{P}(\vec{\alpha})}{\partial t}=\frac{1}{2\tau }%
\sum_{\lambda =1}^{N}\left( \frac{\partial \left( \alpha _{\lambda }\mathcal{%
P}(\vec{\alpha})\right) }{\partial \alpha _{\lambda }}+\frac{\partial \left(
\beta _{\lambda }\mathcal{P}(\vec{\alpha})\right) }{\partial \beta _{\lambda
}}\right)  \\
+\frac{V}{2}\sum_{\lambda =1}^{N-1}\left\{ -\alpha _{\lambda +1}\beta
_{\lambda +1}\left( \frac{\partial \left( \alpha _{\lambda }\mathcal{P}(\vec{%
\alpha})\right) }{\partial \alpha _{\lambda }}+\frac{\partial \left( \beta
_{\lambda }\mathcal{P}(\vec{\alpha})\right) }{\partial \beta _{\lambda }}%
\right) \right.  \\
+(\alpha _{\lambda }\beta _{\lambda }+1)\left( \frac{\partial \left( \alpha
_{\lambda +1}\mathcal{P}(\vec{\alpha})\right) }{\partial \alpha _{\lambda +1}%
}+\frac{\partial \left( \beta _{\lambda +1}\mathcal{P}(\vec{\alpha})\right)
}{\partial \beta _{\lambda +1}}\right)  \\
+2\alpha _{\lambda +1}\beta _{\lambda +1}\frac{\partial ^{2}\mathcal{P}(\vec{%
\alpha})}{\partial \alpha _{\lambda }\partial \beta _{\lambda }} \\
\left. -\frac{\partial ^{2}\left( \alpha _{\lambda +1}\alpha _{\lambda }%
\mathcal{P}(\vec{\alpha})\right) }{\partial \alpha _{\lambda +1}\partial
\alpha _{\lambda }}-\frac{\partial ^{2}\left( \beta _{\lambda +1}\beta
_{\lambda }\mathcal{P}(\vec{\alpha})\right) }{\partial \beta _{\lambda
+1}\partial \beta _{\lambda }}\right\} \,.  \label{eq:FP}
\end{multline}%
According to the Ito calculus, a Fokker-Planck equation of the form
\begin{equation}
\frac{\partial \mathcal{P}(\vec{\alpha})}{\partial t}=-\sum_{n}\frac{%
\partial \left( f_{n}(\vec{\alpha})\mathcal{P}(\vec{\alpha})\right) }{%
\partial \vec{\alpha}_{n}}+\sum_{nm}\frac{\partial ^{2}\left( M_{nm}(\vec{%
\alpha})\mathcal{P}(\vec{\alpha})\right) }{\partial \vec{\alpha}_{n}\partial
\vec{\alpha}_{m}}  \label{eq:FPmatrix}
\end{equation}%
with the symmetric matrix $\mathbf{M}(\vec{\alpha})=\textstyle{\frac{1}{2}}%
\mathbf{B}(\vec{\alpha})\mathbf{B}(\vec{\alpha})^{\mathrm{T}}$ is equivalent
to the set of Langevin equations~\cite{Drummond1980}
\begin{equation}
\frac{\partial \vec{\alpha}_{n}}{\partial t}=f_{n}(\vec{\alpha})+B(\vec{%
\alpha})_{nm}\eta _{m}
\end{equation}%
where $\eta _{m}$ are independent stochastic Gaussian noise terms, defined
by $\langle \eta _{m}(t)\eta _{n}(t^{\prime })\rangle =\delta _{mn}\delta
(t-t^{\prime })$. Observable quantities are obtained from averaging the
Langevin equation over a stochastically generated ensemble. In particular,
the average occupations $\langle \hat{a}_{\lambda }^{\dagger }\hat{a}%
_{\lambda }^{\vphantom{\dagger}}\rangle =\langle n_{\lambda }\rangle $ and
second order correlations $\langle \hat{a}_{\lambda }^{\dagger }\hat{a}%
_{\lambda }^{\dagger }\hat{a}_{\lambda }^{\vphantom{\dagger}}\hat{a}%
_{\lambda }^{\vphantom{\dagger}}\rangle $ are given by:
\begin{subequations}
\label{eq:correlators}
\begin{align}
\langle \hat{a}_{\lambda }^{\dagger }\hat{a}_{\lambda }^{\vphantom{\dagger}%
}\rangle & =\langle \beta _{\lambda }\alpha _{\lambda }\rangle \,, \\
\langle \hat{a}_{\lambda }^{\dagger }\hat{a}_{\lambda }^{\dagger }\hat{a}%
_{\lambda }^{\vphantom{\dagger}}\hat{a}_{\lambda }^{\vphantom{\dagger}%
}\rangle & =\langle \beta _{\lambda }\beta _{\lambda }\alpha _{\lambda
}\alpha _{\lambda }\rangle \,.
\end{align}%
The normalized second order correlation function is then defined as $%
g_{2,\lambda \lambda }=\langle \hat{a}_{\lambda }^{\dagger }\hat{a}_{\lambda
}^{\dagger }\hat{a}_{\lambda }^{\vphantom{\dagger}}\hat{a}_{\lambda }^{%
\vphantom{\dagger}}\rangle /\langle n_{\lambda }\rangle ^{2}$. By solving
Eq.~(\ref{eq:FP}) numerically, we are able to study the quantum statistical
properties of bosonic cascades under conditions of macroscopic occupations,
with several millions of particles.

\textit{Two-Level Case.---} Considering the simplest case of two levels, we
obtain the time dependence of the average populations and second order
correlation functions shown in Figs.~\ref{fig:2Levels}a and b, respectively,
for the case where the upper state is initially a coherent state populated
with $\langle n_2 \rangle=5\times10^7$ particles and the ground state is in
the vacuum.
\begin{figure}[h]
\includegraphics[width=\linewidth]{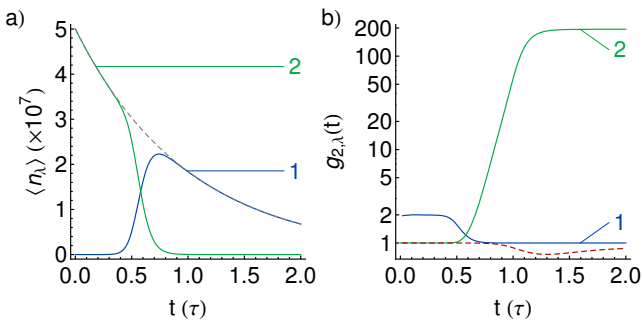}
\caption{Evolution of two bosonic levels following a coherent pulsed
  excitation. a) Average level occupations, $\langle
  n_\protect\lambda%
  (t)\rangle$. The dashed curve shows the exponential decay of the
  total population. b) Second-order correlation function,
  $g_{2,\protect\lambda}(t)$%
  . The dashed curve shows the cross-correlation function
  $\langle\hat{a}%
  ^\dagger_1\hat{a}^\dagger_2\hat{a}_1\hat{a}_2\rangle/(\langle n_1
  \rangle\langle n_2 \rangle)$. A coherent state of $\langle n_2
  \rangle=5\times10^7$ particles is assumed in the highest level and
  we set $V%
  \protect\tau=8.3\times10^{-7}$~\protect\cite{Liew2013}.}
  \label{fig:2Levels}
\end{figure}
The dynamics of relaxation can be well understood: the population from the
upper level decreases in time due to radiative emission and transfers to the
lower level. At early times, the upper state populates the ground state in a
thermal (or chaotic) quantum state, since it merely provides an incoherent
input that increases the average population without developing any other
independent observable~\cite{delValle2011}. Consequently, the particle
statistics of the lower level is initially $g_{2,\lambda=1}(t)=2$ which is
the well known second-order correlation for an incoherent gas of bosons, due
to their indistinguishability. Then, as population increases, stimulated
emission sets in and the upper state now empties much faster and
predominantly into the ground state. This results in the rapid growth of the
ground state population until the upper state is so much depleted that it
cannot compensate for the ground state's radiative losses, at which stage
the ground state starts to decay through its own radiative channel. In this
buildup phase of the ground state, coherence also grows, as seen in the
transition from $g_{2,\lambda=1}(t)=2$ to $g_{2,\lambda=1}(t)=1$. Note that
the state remains coherent from there on as radiative decay alone does not
dephase the condensate. More striking is the statistics of the upper state.
While its second-order correlation function was initially unity, as befits a
coherent state, and remained essentially unaffected in the first phase of
radiative decay and spontaneous emission into the ground state, there is a
pronounced super-bunching $g_{2,\lambda=2}(t)\gg2$ that forms when the state
is rapidly emptied. Admirably, just as the statistics of the ground state
remains equal to one until complete evaporation, the upper state's
statistics also remains pinned at the value it reached when it completed its
transfer. This well defined and high value of $g_2$ for a state
asymptotically approaching the vacuum is due to the mathematical limit of
the two vanishing quantities in Eqs.~(\ref{eq:correlators}) that exists even
for arbitrarily small probabilities of occupation. In practice, numerical
simulations are less stable in this region and experimental measurements
would also be increasingly difficult. Nevertheless, the most interesting
phenomenology is the superbunching that develops in the phase where the
upper state quickly empties into the ground state. In this process, where
one condensate is sucked by another one, the photons emitted radiatively by
the upper state will indeed exhibit a physically observable super-bunched
statistics.

The reason for this peculiar behaviour is to be found in the mechanism of
coherence buildup. Equation~(\ref{eq:master}) is equivalent to a quantum
Boltzmann master equation~\cite{gardiner97a,laussy04a}, in which formalism
coherence---as measured by Glauber's correlation functions---builds up
thanks to population correlations developed by the dynamics, even in the
absence of interactions or external potentials. In the conventional case of
Bose condensation, the Poisson fluctuations of a single condensed state
(usually the ground state), which leads to $g_{n}=1$ for all~$n$, are due to
this single state adquiring the fluctuations of a macroscopic system~\cite%
{laussy12a}. Since the central limit theorem states that the distribution of
a large number of random variables (the excited states) is a Poisson
distribution, so does also fluctuate the condensate in one single state (the
ground state). Each excited state taken in isolation contributes very little
to the condensate in a macroscopic system and its statistical properties are
not significantly altered. In bosonic cascades, however, the asymmetry
between ground and excited states is lost since there can be a few
macroscopically occupied excited states. This is a peculiar configuration
where coherence is adquired from another single coherent state, rather than
from a macroscopic ensemble of weakly occupied thermal states (the
statistical properties of which do not matter, still following the central
limit theorem, as long as they obey generic conditions of independence and
normalization). This peculiarity is the reason why the excited state
develops such a pronounced superbunching when acting as a reservoir for
another condensate to grow in another single state. The fast loss of
coherence from a single state to provide for the coherence of another single
state, results in the superbunching, or extreme chaos, for the provider that
substitutes a macroscopic environment.

Further consequences of this mechanism are that the superbunching should
become stronger for bigger initial populations and weaker for more states in
the cascade. This is expected from the greater dissimilitude of the bosonic
cascade from a single condensate and a macroscopic reservoir: the more
levels there are and/or the less they are occupied, the more the system
resembles the usual scenario. Indeed, numerical simulations show that this
is the case. The dependence of the second order correlation functions,
together with the occupation numbers, on the initial occupation number is
illustrated in Fig.~\ref{fig:2LevelsPowerDependence} and confirms the
expected behaviour of $g_{2,22}$ with higher populations.
\begin{figure}[t]
\includegraphics[width=\linewidth]{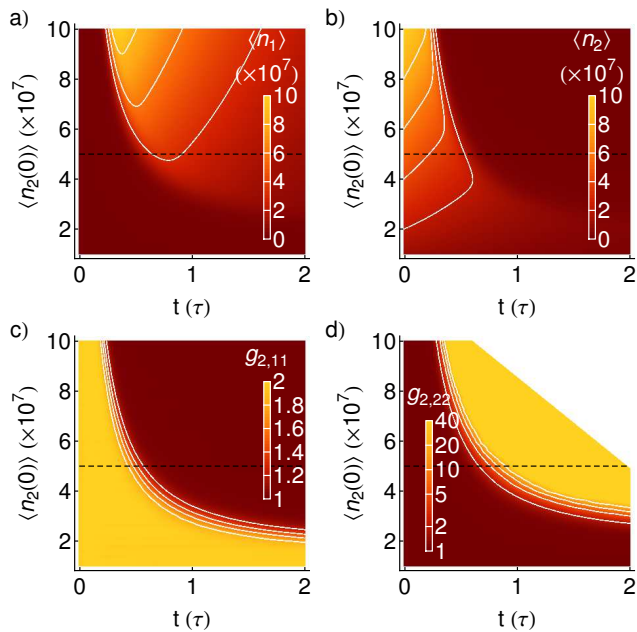}
\caption{Power dependence of the two-level cascade. a) Average occupation of
the lowest level. b) Average occupation of the highest level. c) $g_{2,11}$
of the lowest level. d) $g_{2,22}$ of the highest level. Contours correspond
to the values shown in the colour bars. The horizontal dotted line
corresponds to the initial occupation considered in Fig.~\protect\ref%
{fig:2Levels}. In (d) the value of $g_{2,22}$ is not defined when $\langle
n_2\rangle$ is small.}
\label{fig:2LevelsPowerDependence}
\end{figure}
The lowest mode maintains the behaviour of being initially thermal, but then
becomes a coherent state upon being highly populated, as the upper state
makes the transition from coherent to super-bunched light. The greater and
sooner the super-bunching, the higher the initial population. The effect of
adding more levels in the cascades also results in weaker super-bunching, as
expected, but as it also comes with notable features of its own, we will
discuss it separately.

\textit{Multi-Level Case.---} We now consider a larger cascade composed of
five levels, still with a macroscopic occupation of the highest excited
state ($\langle n_{5}(t=0)\rangle =10^{8}$) and all others empty. The
results are shown in Fig.~\ref{fig:5Levels}.
\begin{figure}[t]
\includegraphics[width=\linewidth]{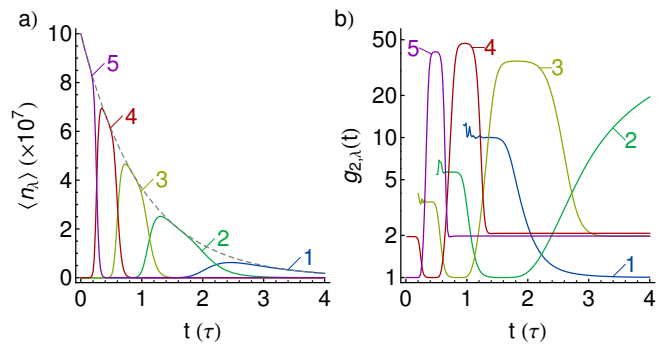}
\caption{Evolution of five bosonic levels following a coherent pulsed
excitation. a) Average level occupations, $\langle n_{\protect\lambda %
}(t)\rangle $. The dashed curve shows the exponential decay of the total
population. b) Second-order correlation function, $g_{2,\protect\lambda }(t)$%
.}
\label{fig:5Levels}
\end{figure}
Looking first at the populations (Fig.~\ref{fig:5Levels}a), we see that
there is a staggered oscillation in the successive levels, corresponding to
the transfers of particles down through each level of the cascade. The
beginning of the relaxation, between the two-highest excited states, is
similar to the case already studied, but since the recipient for the highest
excited state is also the source for another state below, the dynamics is
echoed down the ladder until it reaches the final ground state. In the
process, the dynamics slows down and is tamed in intensity, as particles are
constantly lost to the environment. The second order correlation functions
also exhibit a staggering effect, which can be understood as a
generalization of the behaviour observed in Fig.~\ref{fig:2Levels} for the
two-level case, but with a richer phenomenology. The super-bunching is still
observed when the feeding condensate is sucked into the growing one but this
process gets disturbed when the latter condensate becomes in its turn the
feeding singly-occupied macroscopic state that trades its coherence to the
state below it. This releases the drain on the former condensate, which
therefore relaxes the super-bunching of what is left of the particles there.
In the phase of condensation, the state that grows its coherence does so not
only at the expense of its exciting state, but also of the state below. This
results in a smaller plateau of super-bunching for the latter state that
sandwiches the condensate together with the plateau of large super-bunching
from the exciting state. As a result, the particle statistics for each state
is an intricate sequence of several plateaus joined by abrupt jumps as the
condensate starts to form or starts to empty. For a generic state---that is,
one that is not too close from the most excited state or from the ground
state---the statistical relief has five plateaus: i) a thermal state
starting with, and growing from, the vacuum before the cascade is started;
ii) a plateau of \textquotedblleft small super-bunching\textquotedblright\
as it provides coherence from below to its exciting state; iii) a coherent
plateau as the state is building its own coherence from the excited state
now in its process of avalanche; iv) a plateau of \textquotedblleft big
superbunching\textquotedblright\ as the state feeds the state below,
becoming super-chaotic in the transfer of coherence; finally, v) a thermal
state as the process got transported to states below, with the next state
now in its phase of \textquotedblleft big superbunching\textquotedblright\
and that two states below in that of building its coherence. The most
relevant plateau is that of stage iv), as it is present for all the excited
states and has the stronger signal. All the excited states except that
immediately above the ground state ultimately revert to a thermal state,
once the peculiar dynamics of bosonic cascade is long gone. The complex
concordance of these several stages in statistics with the corresponding
ones in populations can be observed in Fig.~\ref{fig:5Levels}b, showing the
beautiful and peculiar dynamics of quantum statistics in bosonic cascades.

\textit{Conclusion.---} By solving the Fokker-Planck equation for the
positive-P representation of a ladder of bosonic states coupled as
nearest-neighbors, we could compute the quantum statistics of a bosonic
cascade with several levels and for macroscopic occupations. We have shown
how the peculiar nature of this system, where coherence is exchanged between
single states, results in a super-bunching in the phase where one condensate
empties suddenly into the state below. The observation of this strong and
striking quantum optical effect could be made in wide parabolic quantum
wells where excitons are confined as whole particles. It will not only
illustrate fundamental features of coherence buildup in non-interacting
bosonic gases, but also provide new venues for imaging applications, in
addition to the prospects of implementations in the terahertz bandwidth
already offered by these systems.

%\bibliography{Sci}

AK thanks the EPSRC established career fellowship for support.  IAS
thanks ITN NOTEDEV for support. ASS acknowledges RFBR project
15-02-01060 and FPL the project POLAFLOW. SDL is a Royal Society
Research fellow.

\end{subequations}

\end{document}